\documentclass[12pt]{article}
\usepackage[dvips]{graphicx}

\usepackage{color}
\usepackage[x11names,rgb]{xcolor}

\usepackage{tikz}
\usetikzlibrary{arrows,snakes,backgrounds}

\usepackage{amsfonts}
\usepackage{amsmath}
\usepackage{amssymb}
\usepackage{amsthm}
\usepackage{enumerate}

\usepackage[pdftex]{hyperref}
\usepackage{natbib}
\bibpunct{(}{)}{;}{a}{,}{,}
\hypersetup{colorlinks,%
            citecolor=blue,%
            filecolor=blue,%
            linkcolor=blue,%
            urlcolor=blue,%
            }

\usepackage{fancybox}

\renewcommand{\vec}[1]{\mathbf{#1}}

\theoremstyle{remark}

{\begin{Sbox}\begin{minipage}}%
{\end{minipage}\end{Sbox}\fbox{\TheSbox}}

\begin{document}

\title{Stochastic network formation and homophily\\
{\small forthcoming in the \emph{The Oxford Handbook on the Economics of Networks} }}

\author{Paolo Pin\thanks{%
Dipartimento di Economia Politica e Statistica, Universit\`a degli Studi di Siena, Italy.
Email: paolo.pin@unisi.it.}
 \and 
Brian W. Rogers\thanks{%
Department of Economics, Washington University in St. Louis.
Email: brogers@wustl.edu}}
\date{May 2015}

\maketitle


\bigskip


%
%
%
%
%
%
%
%
%

\newpage

\section{Introduction}
\label{sec:introduction}

Many aspects of our lives are affected by social interactions with others.  Often, the way these relationships operate is influenced by interactions between others.  For example, an opinion communicated by one person to another is likely to have been influenced by the outcomes of previous conversations between different pairs of individuals.  That is, a given relationship typically does not live in isolation, but rather is embedded in a social structure consisting of many relationships.  It is now well-recognized that, accordingly, if one wants to understand many kinds of social phenomena, it is imperative to study the \emph{networks} of these social relationships, rather than restricting the level of the analysis to any particular relationship.  
It thus becomes important to understand the structure of social networks, how and why they form, and how their properties relate to the interactions, behaviors, and processes of diffusion occurring through them.  

So it is not surprising that researchers have employed a vast  range of methodologies towards these ends.  
One characteristic along which to broadly categorize studies of network formation is in terms of those based primarily on strategic motivations as opposed to those based primarily on random events.  
Naturally, real networks are formed through a combination of randomness and self-motivated behaviors.  
This observation is reflected in modeling choices and, indeed, there are numerous analyses that, to varying extents, incorporate both random and strategic considerations into modeling link formation processes.  Nevertheless, the literature can largely be sorted according to whether strategic considerations or, instead, random events, constitute the main guiding force behind which links are formed and which links are not formed.  
We here take up the latter category, and aim to survey and discuss the literature that studies network formation by viewing random events as the essential ingredients.  See in this handbook \cite{VMChapter5} for a discussion of strategic-based models, and \cite{FernandoChapter7} for a survey of the literature that explicitly takes into account the coevolution of networks and behaviors.  


We posit that random network models form the foundation of modern network theory.  The literature has demonstrated  that random network models are far superior to purely strategic models in terms of generating networks with properties that are consistent with field data on the structure of large social networks.  This conclusion has several derivative implications.  First, a number of random network models offer tractable frameworks within which to study incentives, and where it is possible to draw conclusions about how network structure impacts behavior.  Such an approach can either allow strategic considerations to be brought into the analysis of link formation or, quite distinctly, it can allow the researcher to study, e.g., diffusion processes or opinion formation from a strategic perspective in a network context.  This kind of work is particularly important since, without modeling the incentives of agents in the network, it is not possible to discuss, e.g., which network structures perform better than others from a social perspective, nor is it possible to provide policy-oriented conclusions.  

Second, empirical work on network formation can be based only on random network models.  As data becomes increasingly available along with the computing power to analyze it, empirical work will continue the trend of becoming increasingly important and useful.  
It is worth nothing that random network models constitute the benchmark with respect to which network properties are often described.  That is, recognizing that real networks are not purely random, once we understand the macro features of purely random networks, we can begin to understand their discrepancies from real networks as attributable to non-random behavior.  
Always, e.g., when we refer to \emph{small world networks}, to \emph{skewed degree distribution} or to \emph{high clustering} in a specific network, we make an implicit comparison between the particular network we refer to and the expected outcome of some random network generation process.

\bigskip

By definition, the work we survey models the connections that form between agents as probabilistic events.  The simplest such models, which serve as a baseline for the rest of the literature, view these events as independent and identically distributed across pairs of agents.  Implicit in such an approach, of course, is that agents have no observable characteristics that are relevant to the formation of the network through affecting the probabilities with which different agents connect to each other.  A more descriptive or realistic approach would have to accommodate the possibility of various kinds of correlations in linking outcomes.  In essence, the likelihood with which an analyst views a connection between a given pair of agents to be formed is influenced by what is known about the characteristics of the agents, and possibly also by the presence or absence of other links.  Agents' characteristics will influence link outcomes if, for example, there are complementarities or substitutability of the characteristics in the context of the relationship.  

In this regard, the single most important force relating to agents' characteristics is \emph{homophily}, which is the tendency of agents with similar characteristics to link with each other.  Because homophily is so robustly observed across many contexts and dimensions, models that incorporate (the possibility of) homophily are central to the study of random networks in general, and often times it is the presence of homophily that allows a model to capture real-world properties that are missing from baseline models.  Homophily has been documented by sociologists along a multitude of dimensions in self--reported friendship networks, and it is present in nearly all social networks if one examines the relevant characteristics.
It has also been shown that homophily plays an important role in governing the outcomes of many important network-based phenomena, through studying models of na\"ive learning \citep{golub2012homophily}, 
population games (\citealt{dalmazzo2014communities}), strategic network formation games (\citealt{jackson2005economics} and \citealt{de2009ethnic}),  and also random network processes (\citealt{currarini2009economic,currarini2010identifying}, \citealt{bramoulle2012homophily} and structural models for empirical estimation that we discuss in Section \ref{sec:structural}).  With modern communication technologies, the costs of maintaining geographically separated connections has been dramatically reduced, allowing for different patterns of social interactions.  In particular, homophily is increasing along a number of prominent dimensions (see, e.g.,  \citealt{rosenblat2004getting}).

\bigskip

%

%

Often it is difficult to disentangle from real data the role of homophily from that of peer effects and contagion (on this see, among others, \citealt{ozgur2013dynamic}, \citealt{mele2010structural} and \citealt{goldsmith2013social} -- and \citealt{ChandrasekharChapter21} in this book).
Actually, the following econometric problem is a main open issue: what are the conditions on the data for which this identification is possible (\citealt{angrist2014perils}, \citealt{shalizi2011homophily})?
We will see that most current answers to this question are indeed based on specific random networks models.

%
%
%
%
%

\bigskip

Our presentation below is delineated by a classification of models according to how they treat population dynamics.  In particular, we first distinguish network formation models that are ``one shot'' from those that are dynamic and, in the latter case, according to  whether the number of nodes varies over time, and how.
In particular, and also following the chronological order of how these models came into the economics literature, Section \ref{sec:oneshot} illustrates one shot models with a fixed number of nodes,
Section \ref{sec:growing} presents growing random network models,
and Section \ref{sec:steady} discuss dynamical models in which the number of nodes is fixed or approaches a steady state.
In Section \ref{sec:homophily} we discuss the economic literature on homophily and its deep relation with the one on random networks.
Finally, Section \ref{sec:conclusion} concludes.\footnote{We remark that many of our topics have received coverage in the books of \cite{Fernando07} and \cite{Jackson_book}.
Relative to those contributions, we try here to convey a unified view, that includes also the more recent developments, especially in the fields of network games and empirical estimation models.}

\section{One shot models with a fixed population}
\label{sec:oneshot}


There are $2^{n (n-1)}$ directed unweighted networks between $n$ nodes, and $2^{\frac{n (n-1)}{2} }$ undirected ones.
A one-shot network formation model is a probability distribution $P_n$ among this finite (but large) number of networks.\footnote{%
The recent paper of \cite{acemoglu2013network} uses exactly this general definition.}
We distinguish in this section between those models that consider each link as an independent event, those that impose some of the network characteristics, and finally those that assume a probability distribution over the characteristics, assuming a correlation in the probability of link occurrence.  

\subsection{Independent link formation: a baseline}

The \cite{erdos1960evolution} model is at the foundation of modern network theory.
It is a simple model in which, starting from $n$ nodes, each of the $\frac{n(n-1)}{2}$ potential undirected links is randomly formed according to an i.i.d.~probability $p(n)$.  
This model presents remarkable \emph{threshold properties}: i.e.~topological characteristic that arise almost surely if and only if $p(n)$ is above a certain threshold as $n$ grows without bound.
Informally, we say that a network property $\mathcal{A}$ is \emph{monotonic} if, given that it holds a.s. for a certain $p(n)$, it also holds for any $\hat{p}(n)>p(n)$, i.e., it continues to hold when adding links to the network.
So, for example, the property of being path-connected, or having at least one cycle, are monotonic properties, while the property of there existing isolated nodes is not.\footnote{Path-connected means that between every pair of nodes there exists a sequence of links connecting them.  A cycle is a path that terminates at the same node where it originates.}
Then we define as $\mathcal{P}_{n,p(n)} (\mathcal{A})$ the probability that the outcome of the i.i.d.~generation process on $n$ nodes, with probability $p(n)$, satisfy property $\mathcal{A}$. 
The authors show that for each monotonic property $\mathcal{A}$ there exists a discrete function $t(n)\in [0,1]$, which is also a probability depending on $n$, such that if $p(n)$ grows asymptotically larger than $t(n)$ then at the limit $n\rightarrow \infty$ we have $\mathcal{P}_{n,p} \rightarrow 1$, if instead $p(n)$ grows asymptotically smaller  than $t(n)$  then at the limit $n\rightarrow \infty$ we have $\mathcal{P}_{n,p} \rightarrow 0$ (while there may be an interior outcome when $\lim_{n \rightarrow \infty} \frac{p(n)}{t(n)}$ is a positive constant).

To provide some examples,  $t(n)=1/n$ is the threshold above which the network has almost surely at least one cycle and a unique giant component, while $t(n)=\log(n) /n$  is the threshold above which the network is almost surely connected.\footnote{A component is a maximal set of connected nodes.  A giant component is a component that contains a non-trivial proportion of nodes as $n$ grows.}$^,$\footnote{%
All this is treated extensively, with more examples, also in \cite{bollobas1981diameter,bollobas1998random} and in \cite{Jackson_book}.}
Note that a specific monotonic property that we consider can also have clear economic implications: so, for example, \cite{elliott2013financial}, in a model on financial contagion, show that contagion on random networks display the same threshold properties.

\subsection{Random graphs of a given degree distribution}


A generalization on the \cite{erdos1960evolution} model is given when some of the characteristics of the network are fixed.
In this sense $P_n (g)$ is positive only if $g$ satisfies those characteristics.
In principle it is possible to fix many of the characteristics, such as the clustering coefficient or the diameter.\footnote{There are different ways to measure clustering, but the essential idea is to measure the frequency with which two nodes who have a common neighbor are connected.  The distance between a pair of nodes is the number of links in the shortest path connecting them.  A graph's diameter is the maximum distance between a pair of nodes.
The degree of a node is the number of links it has.  A degree distribution describes the frequencies of different degrees in the population.  The chapter of \cite{ChandrasekharChapter21} defines these concepts, and others that we discuss, more formally. }
As an example \cite{bianconi2008entropy} studies ``network ensembles with the same degree distribution, the same degree correlations and the same community structure of any given real network'' (on this see also \citealt{bianconi2008entropies} and the discussion on exponential random graph models in Section \ref{sec:markov_graphs} below). 
There is also a category of models based on an exogenous differentiation between the nodes, which determines heterogeneous probabilities of linking (we discuss this in Section \ref{sec:blockmodels}).

However, most of the literature, and particularly in theoretical economics, has focused on networks with a predetermined degree distribution.
These are called \emph{configuration models} (\citealt{bender1978asymptotic}, \citealt{molloy1995critical,molloy1998size} and \citealt{chung2002average}):
given a set of nodes $n$, and an $n$--dimensional vector $\vec{d} \in \{1,2,\dots,n-1\}^{n}$ for the degree distribution, we attribute positive probability only to those networks with that degree distribution. \\
If the network that we want is directed, then there is a natural way to define the probabilities: we simply assign to node $i$ a uniformly random subset of size $d_i$ from the other nodes. 
This provides, with the help of combinatorics, an unambiguous way to compute $P_n$ that will determine i.i.d.~uniform expectations for each node on her neighbors in the realized networks. \\
If instead we attribute positive probabilities only to undirected networks then  we accept only those outcomes where node $i$ sends a link to node $j$ if and only if node $j$ reciprocates. 
Recent works like \cite{karrer2011stochastic} show that in this case things can be complicated: should we attribute uniform probabilities to each of the admissible network outcomes, or should we instead try to maintain uniform expectations for each node on her neighbors in the realized networks?\footnote{%
It is possible to show with a very simple example that, especially when $n$ is low, the two considerations are incompatible: there is only one undirected network with $n=4$ and $\vec{d}=(1,2,2,3)$, and in this network node $1$, who has degree $1$, can only be matched with node $4$, who has degree $3$.}

\bigskip


The configuration model has found extensive application in a recent literature on Bayesian network games, especially since \cite{galeotti2010network}.
In this literature it is assumed that agents (nodes in the network) know the degree distribution, their own degree and, as a consequence, the probability distribution of their neighbors.\footnote{Papers that have anticipated some of the theoretical results are \cite{jackson2007diffusion}, \cite{sundararajan2007local}, \cite{lopez2006contagion,lopez2008spread} and \cite{galeotti2011complex}. See also \cite{BKChapter8} in this book.} 
In this framework, one can consider the case of no correlation in the degrees of linked nodes, or it could be the case instead that, e.g. because of \emph{assortative degree correlation} (see also Section \ref{sec:growing} below on \emph{assortativity}), nodes with high degree expect to meet other high degree nodes with higher probability than in the case of no correlation.\footnote{%
On the cases with degree correlation, see the discussion in \cite{galeotti2010network} and the general framework proposed in \cite{feri_pin2014}.}
Then every agent chooses, before the network is realized, an action to be played in a given game on the realized network, computing in the Bayesian way the expected action profile of her neighbors.\footnote{In this literature, applications to games of vaccination against the risk of contagion are in \cite{galeotti2013diffusion,galeotti2013strategic} and in \cite{goyal2014adaptive}; an application to public goods is in \cite{lopez2013public}; applications to peer effects and influence are in \cite{lopez2012influence} and \cite{jackson2013diffusion}.
Also some market models are built on this framework: \cite{nermuth2013informational} provide an application to a price competition between firms, in a bipartite network of  firms and consumers; \cite{fainmesser2013value} consider instead a game of purchasing choice in a network of consumers with positive complementary peer effects for the consumption of a specific good.
Note finally that in support of this approach, \cite{charness2012equilibrium} provides experimental evidence of the results from \cite{galeotti2010network}.}

\subsection{Incorporating other network statistics}
\label{sec:markov_graphs}


It is often the case in real world social networks, that if node $i$ has two \emph{friends} $j$ and $k$, then it is likely that also $j$ and $k$ are \emph{friends} together.
This property is known as \emph{clustering}; it was first considered and defined in the literature on complex networks by \cite{watts1998collective} (and alternative definitions are in \citealt{newman2003structure}) and was then empirically measured by \cite{ravasz2003hierarchical} and \cite{vazquez2003growing}.
However, the models that we discuss below, dealing with this issue, date back to the 1970s and '80s.

\bigskip

The main idea is that a network exhibits clustering if it shows a considerable amount of closed triangles (which would be unlikely to appear in the \citealt{erdos1960evolution} model). 
So, even if the classical intuition of why clustering arises is dynamical (and we will discuss in Section \ref{sec:growing} models reflecting such an intuition), one could imagine a one--shot network formation model in which triangles are assigned with some specific probability.
This is what is essentially done in the class of \emph{Exponential Random Graph Models} (ERGM, also called $p^*$ models -- see \citealt{frank1986markov}, \citealt{strauss1986general}, \citealt{wasserman1996logit}, \citealt{anderson1999p}, \citealt{park2004solution,park2005solution} -- see also \citealt{besag1974spatial} for a similar concept in the field of spatial econometrics).  Since Section 4 in the chapter of \cite{ChandrasekharChapter21} in this handbook presents a detailed treatment of ERGMs, we refer the reader there for details, and we keep our discussion here brief.

\bigskip

Formally, the analyst specifies a set of simple statistics that are to be satisfied by the network $g$, such as the number of links $L(g)$, the number of closed triangles $T(g)$, and possibly a number of more complicated structures such as loops of a certain length or clusters of $k$ nodes, obtaining, overall, $\ell$ structures of interest, denoted by $(\alpha_1(g)$, $\alpha_2(g)$, \dots, $\alpha_{\ell}(g))$. \\
Now, the probability of a network $g$ that satisfies exactly the $\ell$-dimensional vector $\vec{\alpha}$ of natural numbers, one for each of the statistics, will be given by a probability
\begin{eqnarray}
\label{eq:ergm}
P_n (g) & \propto & \exp \Big( \beta_1 \alpha_1 (g) +  \beta_2 \alpha_2 (g) + \dots + \beta_{\ell} \alpha_{\ell} (g) \Big) \nonumber \\
& = & \frac{\exp \Big( \beta_1 \alpha_1 (g) +  \beta_2 \alpha_2 (g) + \dots + \beta_{\ell} \alpha_{\ell} (g) \Big)}{
\sum_{g' \in \mathcal{G}} \exp \Big( \beta_1 \alpha_1 (g') +  \beta_2 \alpha_2 (g') + \dots + \beta_{\ell} \alpha_{\ell} (g') \Big)} \nonumber \\
& = & \exp \Big( \beta_1 \alpha_1 (g) +  \beta_2 \alpha_2 (g) + \dots + \beta_{\ell} \alpha_{\ell} (g) - \kappa \Big) \ \ .
\end{eqnarray}
where $\kappa$ is a common constant for the model. \\

%
%

An important issue is that calibrating an ERGM model on real data becomes computationally impossible, due to the need to estimate the value of the constant or, equivalently, from equation (\ref{eq:ergm}), of the quantity 
\[
\sum_{g' \in \mathcal{G}} \exp \Big( \beta_1 \alpha_1 (g') +  \beta_2 \alpha_2 (g') + \dots + \beta_{\ell} \alpha_{\ell} (g') \Big) \ \ .
\]
All available approximation techniques (see \citealt{bhamidi2008mixing} and \citealt{chatterjee2013estimating}) rely on the assumption of \emph{independent links}, which is against the spirit of the model.\footnote{%
\cite{he2013estimation} try to solve the variational problem of \cite{chatterjee2013estimating} with an approximation similar to mean field.}

\cite{chandrasekhar2012tractable} show that these techniques prove themselves to be inaccurate when tested against simulations.
The reason is, as they show, that if the parameters of the ERGM are all non-negative (with exclusion of the parameter attached to the number of links), then asymptotically the ERGM is indistinguishable from an \cite{erdos1960evolution} graph (or a mixture of them). 
For this reason, \cite{chandrasekhar2012tractable} propose \emph{Statistical Exponential Random Graph Models}, in which, instead of computing exactly equation (\ref{eq:ergm}), one assumes sparsity of the network and approximates for each $\ell$-dimensional vector $\vec{\alpha}$  the log--likelihood $K(\vec{\alpha})$ of obtaining one of the networks satisfying $\vec{\alpha}$ from the random network formation process.
Then, the problem reduces to estimating  
\begin{eqnarray}
P_n \left( g | \alpha_i (g')= \alpha_i \ \forall~i \in \{1,\dots,\ell\} \right) = \frac{K(\vec{\alpha}) \exp \left( \vec{\beta} \cdot \vec{\alpha} \right) }{ \sum_{\vec{\alpha}'} K(\vec{\alpha}') \exp \left( \vec{\beta} \cdot \vec{\alpha}' \right)} \ \ ,
\end{eqnarray}
where it is now possible to compute the sum in the denominator. 
\cite{chandrasekhar2012tractable} successfully  calibrate this method to real social network data for 75 Indian villages with a population each in the  order of hundreds.

\subsection{A simple approach to capture basic properties}
\label{sec:wattsstrogatz}

\cite{granovetter1973strength} is a pioneering paper in sociology, that after having analyzed job contact networks through interviews, proposed the \emph{strength of weak ties} idea.
Essentially, since the people we meet more often are also those that share most of our daily experience, and are exposed to correlated sources of information, it is through the people we meet occasionally (with whom we have the weak ties) that we obtain the most valuable information.
Also, the weak ties make it possible to decrease the overall average distance of a  social network, bridging together communities that would otherwise be disconnected, or at least very far apart.\footnote{%
This is reminiscent of the work of \cite{burt2004structural}, who observes how the nodes who have these long--range weak ties are also the key nodes in the diffusion and aggregation of new ideas.}


With this concept in mind, \cite{watts1998collective} proposed a model where, starting from a fixed lattice (a ring, or a two--dimensional grid) that establishes naturally an Euclidean distance between the nodes, some links are rewired randomly, and this rewiring makes the distances that they bridge in the underlying structure arbitrarily large.  
Historically, this was the first step in the field of \emph{complex networks}, that we will discuss in next Section \ref{sec:growing}.  We refer the interested reader to the chapter of \cite{WattsChapter} for more details and perspective.  It is somewhat surprising that the   \cite{watts1998collective} model has not been used more in the theoretical economic literature.  One potentially interesting application that could prove useful in future research is to the idea of ``searchability'' of a network, also discussed at length in \cite{WattsChapter}.  For example, if one needs to find a particular input for production from a node in the network, one way to do this is to ask one's neighbors, who can ask their neighbors, and so forth.  We have a less than complete understanding of which networks are more easily searchable, and so perhaps more efficient under varying conditions.\footnote{We discuss in Section \ref{sec:blockmodels}
another way to tell the weak ties story that has been adopted in some theoretical economic papers, where the underlying distance is given by a segregation of the nodes in communities, given by homophily.}  One could also consider labor models of job search based on this framework, especially given that  \cite{kleinberg2000navigation,kleinberg2004small,kleinberg2006complex} derive analytical results to study decentralized search and information diffusion in the context of this model.




\section{Network formation with a growing population}
\label{sec:growing}

In the late 1990s, increased availability of data allowed researchers to study the empirics of  \emph{complex networks} at a greater level of detail across a range of applications, including friendship networks, professional networks such as coauthorship, and more recently using social media and other online data such as blogs.  This allowed an analysis of some key aspects of the macroscopic structure of  large social networks for the first time.  As it turns out, many networks share a number of common empirical features.  A good survey on this is \cite{newman2003structure}.  We will briefly discuss three of the most important empirical regularities.  

The first relates to the distribution of connectivity in the network across nodes.  Specifically, starting from the analysis of the world wide web (see, e.g., \citealt{albert1999internet}), and then generalized to other complex networks (by \citealt{baal}),
it was shown that the degree distribution exhibits heavy tails .   That is, there are many nodes with relatively few connections and, perhaps more importantly, there are also many nodes that are highly connected.  The observation of heavy-tailed degree distributions has sometimes been strengthened to claim that networks exhibit a power-law degree distribution, at least in the upper tail.  Intuitively, if one is interested in understanding processes that operate across a network, the degree distribution is of central importance.  For example, if a disease is spreading through a population, the outcome will be heavily influenced by the prevalence of high degree nodes, since they are both more likely to be infected and then more likely to spread the disease to many other nodes.  

The second regularity is high clustering, as discussed in Section \ref{sec:markov_graphs}.  That is, two given nodes who have a common neighbor are more likely to be connected.  Networks with high clustering thus have strong correlations in link patterns, and tend to have highly interconnected subsets of individuals.  The level of clustering is also very important for how certain processes operate on a network.  If one is interested in how information (e.g., about job openings) spreads, a highly clustered network will generate a lot of repeated information.  

The third regularity is positive assortativity, which is to say that highly connected nodes are more likely to have highly connected neighbors than poorly connected nodes (see, e.g, \citealt{newman2002assortative}).  

The fairly robust presence of these properties begs for a common explanation of why they should arise across a wide range of settings.  A rough intuition is the following.  As a network evolves, there is naturally some level of heterogeneity across nodes, arising perhaps partly through chance, and this results in differential numbers of connections across nodes.  Then, as new connections are formed, it is easier or more likely to come into contact with the nodes that already have more connections.  This could be because of their initial popularity (even if it is by chance), or more mechanically if there is some process of searching for nodes along existing links, then nodes with more links are more likely to be found.  The key implication is that any initial differences in connectivity will tend to be exacerbated through the process oof forming additional links.  Thus, eventually, one expects there to be a significant number of very highly connected nodes, since more connections lead to a fast rate of acquiring even more connections.  At the same time, one expects there to be many poorly connected nodes, as having few connections means that new connections are harder to come by.  

If there is a search process that is responsible for the formation of new connections, then that process will also tend to create clustering.  One can imagine meeting the friends of current friends and connecting to them, as in \cite{jackson2007meeting}, or copying the connections of one's neighbors, as in \cite{vazquez2003growing}.  Either of those processes directly induces patterns in which one's friends tend to be friends of each other, precisely because the common friend generated the meeting in the first place.  

Such a search process operates over time, and nodes accumulate new connections gradually.  Thus, a node's degree is typically correlated with its age: it is older nodes who have had more time to accumulate many connections.  So if one focuses on the most highly connected nodes, they tend to be older nodes and, as such, tend to be connected to other old nodes, as those are the ones who were available to receive connections earlier on in the evolution of the network.  

What this intuition suggests is that models that explicitly account for network growth, especially those based on an explicit search process, have the potential to offer a parsimonious explanation for some of the most robust empirical regularities of real networks.  Our focus in this section is on understanding the essential properties of such models.

\subsection{Uniform link formation: a baseline}\label{simpleGrow}

\cite{Jackson_book} presents a simple model of a growing network that we summarize briefly here, as it provides a useful benchmark and introduces some of the techniques used in analyzing such models. \\
Time is discrete and at time $0$ there are $m$ nodes arranged in a fully connected network.
At time $t$ a new node enters (let us call this node also $t$, without ambiguity) and casts with uniform probabilities $m$ links to some of the $t-1+m$ nodes that are currently present in the system.\footnote{%
It is essentially this rule of \emph{attachment}, that here is uniform, that characterizes the different growing network models.
For a general framework for growing networks see the model proposed by \cite{lobel2013preferences,lobel2014information}.}
In this model, if we aggregate out-- and in--degrees, the expected degree of node $i$ at time $t$ is
\begin{equation}
\label{eq:exponential1}
m + \frac{m}{i+1} +  \frac{m}{i+2} +  \dots +  \frac{m}{t} \simeq m \left( 1+ \int_{i}^{t} 1/\tau ~ d \tau \right) = m \left( 1+ \log \left( \frac{t}{i} \right) \right) \ \ .
\end{equation}
Nodes that have expected degree less than $d$ at time $t$ are those such that
\begin{equation}
\label{eq:exponential2}
m(1+log(t/i)) < d  \ \ \Rightarrow \ \ i > t ~ e^{-\frac{d-m}{m}} \ \ .
\end{equation}
From here, it is straightforward to see that the process thus generates an exponential degree distribution.  In a sense, this model can be seen as the analogue of the Erdos-Renyi model for the setting of a growing network, in that all nodes form their sets of links independently and with uniform attachment probabilities.  Notice, in particular, that the probability of acquiring new links from entering nodes is independent of a node's current degree.  As this model is therefore lacking a ``rich get richer'' aspect of link accumulation, it is not surprising that it generates a thin-tailed degree distribution.
Moreover, the model does not generate high clustering.  Notice that as the network grows, it becomes increasingly sparse.  In asking whether two given nodes are connected, conditioning on the event that they have a common friend does not change the answer, and so clustering coefficients tend to zero as the population grows.  
Notice, however, that assortativity does occur.  The highest degree nodes tend to be the oldest nodes, who tend to be connected to each other.  One takeaway is that assortativity is perhaps one of the most basic consequences of a wide range of growing network models.  

\bigskip

Before going on, let us examine the implicit assumptions that we have used to obtain the expression (\ref{eq:exponential2}) which delivers the exponential degree distribution.
We have adopted a continuous time approximation to derive a closed form expression in equation (\ref{eq:exponential1}).
Also, we have taken expected outcomes to compute the degree distribution in (\ref{eq:exponential2}), not considering any other higher order moment of the distribution around this first order approximation.
This simplifying approach is what is called \emph{mean field} approximation, and we will use it also below to consider the more complicated models.  The technique is extremely helpful because the underlying link formation model, while simple at the individual level, generates a highly path-dependent stochastic process that is difficult to analyze directly.  
Some research has derived analytical results without having to invoke a mean field approximation, but more often the literature has proceeded by testing, through simulations, that the outcomes predicted by the mean field approach provide a reasonably good description of the real process.

\subsection{Preferential attachment}

We now turn our attention to models that do a better job of capturing the empirical regularities discussed above.  In particular, the main point of this subsection is that by adjusting the link probabilities away from uniform attachment in a fairly natural way, one can readily generate a heavy-tailed degree distribution.  This idea originated in \cite{price1965networks,price1976general}, who built on \cite{simon1955class}, but he did not see the generality of his model.  The mathematics underlying the approach first appeared in \cite{yule1925growth}.  However, the idea re-entered the literature more recently in  \cite{baal}, at which point the power of the idea became more widely recognized.  

The essential idea is to specify the linking process such that a node's probability of attracting a new link at any given moment of time is proportional to its current degree at that time.  Let us denote node $i$'s degree at time $t$ by $k_i(t)$.  Adjusting the model above to account for preferential link attachment, and applying the mean field approximation, we have 
\[\frac{d k_i(t)}{d t} = m\left(\frac{k_i(t)}{2mt}\right) = \frac{k_i(t)}{2t},\]
since each of the $m$ links formed at time $t$ reaches agent $i$ with probability equal to agent $i$'s degree divided by the sum of all degrees in the network.  Starting from the initial condition $k_i(i)=m$, we reach the  solution 
\[k_i(t) = m\sqrt{t/i}.\]
Thus, the nodes $i$ that have degree $k_i(t)\leq d$ at time $t$, for some given $d$ are such that 
\[ m\sqrt{t/i} < d \Rightarrow i >\left(\frac{m}{d}\right)^2.\]
This produces a degree distribution given by $F(d) = 1-\left(\frac{m}{d}\right)^2$, which is a power-law distribution and, most notably, has a heavy-tailed (scale free) degree distribution.  

Notice that the preferential attachment model retains the assortativity from the first growing model, and generates a power-law degree distribution through a constant rate of entry of new nodes that form links with probabilities in proportion to existing nodes' current degrees.  

However, at least when the model is interpreted such that link formation is independent across nodes, preferential attachment on its own does not predict high clustering.  A second observation is that, while the power-law distribution exhibits heavy tails, it is a very precise prediction, and frequently does not match well observed distributions, except possibly in the upper tail.\footnote{%
Note also that it is practically impossible for all but the largest data sets if a sample of real data actually exhibit a power law distribution (on this see e.g.~\citealt{virkar2014power}). }
We now describe a model that is motivated by these two obervations.

\subsection{Network search: combining uniform and preferential meetings} 
\label{sec:JR}

We will present a simplified version of the network growth model of \cite{jackson2007meeting}. The model is based on the idea that some connections are formed largely due to random or idiosyncratic forces, while other connections are found through existing links.  That is, one of the main ways that nodes find each other is by meeting them through their existing neighbors.  

More specifically, we shall assume that each entering node first forms a given number, $m_R$, of connections to other nodes found uniformly at random, exactly as in the model of Section \ref{simpleGrow}.  Then, from these partners, the node meets a given number $m_N$ of additional nodes who are connected to his original partners.  The idea is that these network-based meetings capture a process of meeting ``friends of friends''.

There is a related idea of vertex copying in the physics literature first studied by \cite{vazquez2003growing}.  In this approach an entering node first finds an existing node at random, and then makes additional links to all of that node's partners.  While it was not fully appreciated at the time, vertex copying produces many of the same insights as the friends of friends model.  

To study the friends of friends model, we will now account for the orientation of links, distinguishing inlinks from outlinks.  Each node forms $m=m_R+m_N$ outlinks at birth (and never forms additional outlinks).  Inlinks, on the other hand, are accumulated from entering nodes over time.  Applying the mean-field analysis as before, the growth of node $i$'s indegree, now denoted $k_i(t)$, is governed by
\begin{equation}
\frac{d k_i(t)}{d t} = \frac{m_R}{t}+m_N\left(\frac{k_i(t)}{t m}\right),
\label{eq:JR}
\end{equation}
where the first term accounts for the possibility to be found at random, and the second term accounts for the possibility that one of $i$'s in-neighbors is found at random, and then $i$ is found though the network.  Using the initial condition that $k_i(i)=0$, one reaches an indegree distribution described by 
\begin{equation}
F(d) = 1-\left(\frac{r m}{d+r m}\right)^{1+r},
\nonumber
\end{equation}
where $r=m_R/m_N$ is the ratio of the number of links formed at random versus through the network.  

This degree distribution is interesting, since it predicts a power-law in the upper tail, but a somewhat thinner lower-tail, with a shape depending on the value of the parameters.  To see this, note that for large degree $d$, i.e.~the upper tail, the distribution is approximately $\bar{F}(d)=1-(rm/d)^{1+r}$.  On the other hand, for lower degrees, the distribution is not well approximated by a power-law.  

As it turns out, more careful empirical analysis often shows that power-law-like degree distributions are mostly found only for the highest degree nodes, and so this class of degree distributions is fairly well-suited to capturing this feature.  Notice finally that the friends of friends meetings (or similarly the vertex copying model) directly generates high clustering: whenever two nodes meet through the network, it is precisely because they have a friend in common.

\section{Dynamic network formation}
\label{sec:steady}

Here we consider a class of dynamic models where the number of nodes is not necessarily growing, and may even be constant.
Most of the modeling with this approach comes from the economic literature, because of necessities that we will go through case by case.
Such models have been analyzed only very recently also in the other disciplines.\footnote{%
The first work from economists considered situations in which not only the network was endogenous, but also some action played by nodes in the network: first papers are \cite{JacksonWatts02} and \cite{goyal2005network}. 
This approach was then broadened in more complex settings in papers outside of economics, in, e.g., \cite{marsili2004rise} ,
\cite{ehrhardt2006phenomenological,ehrhardt2006diffusion} ,
\cite{holme2006nonequilibrium} ,
\cite{gross2008adaptive} ,
and \cite{konig2011network}.
These results are recently coming back into the economic literature with \cite{fosco2010peer} and \cite{konig2012nestedness}.
With a more applied view,  \cite{lee2013markov} have a market model with a fixed number of farsighted agents, while
 \cite{goldsmith2013social} use  a two--period model with endogenous network and actions for empirical estimation.
Models in which agents make choices in an evolving network are treated extensively by \cite{FernandoChapter7} in this book.
}

\subsection{Search and matching}
\label{sec:pools}

One way to model network formation and, if desired, at the same time behavior mediated by that network, is to base meetings on a search model.  This idea has its roots in the seminal work of \cite{diamond1982aggregate,diamond1982wage}, where the focus was on labor markets rather than networks, per se, but the general methodology is well suited to a more network-centric analysis.  

This approach can incorporate both random and strategic network formation elements.  This is probably why it appears only in the economic literature.  We focus here primarily on the random components; see  \cite{VMChapter5} for a specific treatment of strategic factors.

\cite{currarini2009economic} apply such an approach to studying link formation in the presence of types, focusing on the role that type-based preferences play in equilibrium.  Let us summarize how the model works.  Agents enter the system over time and search for partners, whom they meet randomly.  A given agent stays active in the system until the marginal benefit of additional search falls below its marginal cost.  This random meeting process may exhibit biases, that we discuss in  Section \ref{sec:homophily}.

\cite{immorlica2013emergence} adopt a new approach also based on a matching and search framework.  In their model, agents are born into the system over time and initiate a given number of connections to others found uniformly at random.  At each date, agents choose to cooperate or defect and play a prisoners' dilemma with their current partners.  Then, agents can sever any set of connections that they wish.  Some agents are randomly chosen to die, and then at the next date, any agent with unused connections re-initiates new meetings in the subsequent period.  
\cite{immorlica2013emergence} study the extent to which cooperation can be supported in equilibrium.  Under favorable parameters, full cooperation can be supported.  Otherwise, if there is any cooperation, it takes the form of a specific fraction of the population cooperating, while others defect, so that the two behaviors co-exist.  Even though, in their model, agents are anonymous, and are not threatened by a contagion of defecting behavior, cooperation can be supported because of a network effect: in equilibrium, cooperating agents are able to accumulate many profitable relationships, whereas defectors live in relative isolation, losing all of their links after every period.  

\cite{pin2013cooperation} apply a similar approach to studying immigration policy.  In their model, agents enter either as immigrants or as the offspring of citizens.  The main question is to understand how a policy of punishing defecting immigrants impacts equilibrium outcomes, when individuals are motivated by the outcomes of playing prisoners' dilemmas in a society where connections are made and lost according to the model of \cite{immorlica2013emergence}.  The main finding is that, while increasing punishment increases the level of cooperation in society, it also has the consequence of increasing the rate of defection among citizens.  The intuition is that there is a natural level of cooperation in society dictated by the models parameters.  When the cost of defecting is increased for immigrants through punishment, there will be a partial substitution into defecting behavior by citizens.   

\subsection{Structural models}
\label{sec:structural}

One of the most useful notions of equilibrium for strategic network formation games is \emph{pairwise stability} (see \cite{VMChapter5} in this book): in a nutshell, a network is pairwise stable if, $(i)$ when two nodes are linked together, neither of them prefer to delete the link; and $(ii)$ if they are not linked together, it is not the case that both of them prefer to create the link.
One interpretation of this notion is through a matching pool mechanism: in the society, people are randomly and sequentially matched in couples, and every couple can re--consider their link, and create it with mutual agreement, if absent, or delete it if present, if at least one of them wishes to do so.
This implies a Markov process between all possible network configurations, where pairwise stable equilibria are absorbing states (as discussed in \citealt{JacksonWatts02}).

One can introduce some noise into this system, in the form of possible errors made by the agents, or in the form of shocks on their utility functions, so that the Markov process becomes ergodic, i.e.~with no more absorbing states. 
In this case it is possible to compute ergodic probabilities of each network, as if they were the probabilities of a one--shot network formation model, as those discussed in Section \ref{sec:oneshot} above.
If errors are small, pairwise stable networks would still be likely outcomes, but every possible network will have some positive probability to be in place.

There are at least three recent papers that adopt this approach as a structural model for empirical estimation: they are 
\cite{goldsmith2013social}, \cite{mele2010structural} and \cite{graham2014}.
They will be covered also by \cite{ChandrasekharChapter21} in this book, and as they are related to homophily we will come back to their results in Section \ref{sec:homophily}.
Here we just present the main idea, abstracting from the specific models.

Suppose that there is an undirected\footnote{%
\cite{mele2010structural} actually considers  a directed network.} 
network $g_t$ between $n$ nodes at every moment $t$ in discrete time, and that at time $t$ a matching technology provides positive probability to any pair $(i,j)$ of nodes (but just one pair per period) to consider their mutual link.
Node $i$ gets a utility $U_i (g_t \cup \{ ij \})$ from  the network with that link in place (which could be $g_t$ itself, if originally $ij \in g_t$), and a utility $U_i (g_t \backslash \{ ij \})$ from  the network without that link in place (which could also be $g_t$, in the complementary case).
We call $\Delta_{t,j} U_i = U_i (g_t \cup \{ ij \}) - U_i (g_t \backslash \{ ij \})$, and the same holds for node $j$.
Then, $g_{t+1} = g_t \cup \{ ij \}$ if and only if both
\[
\Delta_{t,j} U_i + \eta_i >0 \ \ \mbox{ and } \ \ \Delta_{t,i} U_j + \eta_j >0 \ \ ,
\]
where $\eta_i$ and $\eta_j$ are i.i.d.~random shocks.
Otherwise, $g_{t+1} = g_t \backslash \{ ij \}$.

This Markov process is ergodic, and when the shocks follow a Weibull distribution (as in the assumptions of the standard  logit model) the problem of estimating ergodic probabilities is very similar to the one discussed in Section \ref{sec:markov_graphs} for ERGM models.
In this case, only for some limiting cases, or when a \emph{potential function} exists for the utilities, is it possible to numerically compute these probabilities.

\section{Homophily}
\label{sec:homophily}

When observing a social network from the real world, and even considering our own perceptions of social networks surrounding us, we always find evidence that similar people are more likely to be connected together than if they were different along some dimension: this applies to cultural classification, age, gender, and so on.

\bigskip

Formally, suppose that the population of a society is heterogeneous, which we model by proposing a partition into a finite number of types, which could denote a classification like age--groups, or self reported racial groups. Each type $\theta$  has a frequency in the population of $w_{\theta} \in [0,1]$, so that $\sum_{\theta'} w_{\theta'}=1$.
Now we check the friends of some agent $i$, where $i$ belongs to type $\theta$.
In this set of friends those that are also of type $\theta$ are a proportion $H_i$ of the neighborhood of $i$ that we call the \emph{homophily} of agent $i$.
In the same way we can define the homophily of a whole group $\theta$ considering the aggregate measure
\[
H_{\theta} \equiv \frac{\mbox{average $\#$ of friends of $\theta$'s members, which are also in $\theta$}}{\mbox{average $\#$ of friends of $\theta$'s members}}
\ \ .
\]
The problem is that $H_{\theta}$ is not a relative measure, therefor we normalize it with respect to $w_{\theta} $ through \emph{imbreeding homophiliy}:\footnote{%
This is also known as the \cite{coleman1958relational} index.}
\[
h_{\theta} \equiv \frac{H_{\theta} - w_{\theta}}{1 - w_{\theta}} \ \ ,
\]
that is positive only when there is positive bias in favor of own type, and reaches its maximum at $1$ when all members of $\theta$ have only friends from $\theta$.

So, we say that a type $\theta$ is homophilous if it has $H_{\theta} > w_{\theta}$, or equivalently $h_{\theta}>0$, and we can also compare more or less homophilous types.
This however does not help us in understanding the causes of this observed segregation: is it due to the choices of the individuals in $\theta$, to the choices of the rest of  the population, or to biases in the meeting opportunities, e.g. because a type is correlated with the occupation of individuals? Also the biases in meetings could result as an effect: as \cite{montgomery1991social,montgomery1992job} has shown, if job places are allocated through informal contacts, as in the \cite{granovetter1973strength} weak ties story, then homophilous behavior results in occupational segregation.
Note that we consider the types as exogenous, and we refer to homophily as the tendency of similar people to connect together.
However types could be endogenous, and diffusion processes  (see \citealt{LambersonChapter11} in this book) could cause people that are connected together to become similar.
When both homophily and diffusion are present, the identification of causalities for observed segregation becomes an extremely hard task, as discussed in \cite{ChandrasekharChapter21} in this book.


The classical economic theory was based on the assumption of a \emph{representative agent}, abstracting from heterogeneity.
However, a preliminary attempt to explain the observed homophily in residential segregation with an elegant and general model is in \cite{schelling1971dynamic}.\footnote{%
\cite{schelling1971dynamic} is based on a grid structure with cells that resemble some of the topological properties of networks, referring to neighbors as those cells that are adjacent in the grid.
Agents have a preference for having neighbors of their same type, and can jump to another random location (in a network setting, they would reshuffle all their links) if unsatisfied.
The main result is that even mild preferences for own type can result in dramatic segregation.}
Separately, from the pioneering work of \cite{becker1973theory}, matching theory has focused attention on \emph{assortative matching}, thereby emphasizing one of the sources of observed homophily.\footnote{Note that assortative matching, which refers to an outcome in which types (skill, beauty, productivity, etc.) of paired agents are positively correlated, is quite distinct from the network-based notion of assortativity, which refers instead to a correlation between the degrees of linked nodes.}
Below we discuss the recent contributions based on homophily and the implications of homophily in various economic models.
Then we end with a discussion of the relationships between the models, which are also based on some randomness in the network formation process,  and the empirical estimation of real network datasets.\footnote{%
\cite{jackson2005economics}, 
\cite{de2009ethnic} and \cite{boucher2015structural} consider  network formation games that are deterministic. We do not  cover them in this chapter.}


\subsection{Static link formation}
\label{sec:blockmodels}

The first implication of homophily is that, as soon as there is some heterogeneity in the characteristics of the nodes, not every pair of nodes will have the same probability to be linked together.
The most intuitive way to apply this is generalizing the \cite{erdos1960evolution} model in a way that assigns to each pair $(i,j)$ a linking probability $p(i,j)$ that depends on the characteristics of the nodes.
In a simplified setting, the characteristic of a node is just a type, as in the definition of homophily above, the linking probability depends on the types of the two nodes, and it is typically higher when the two types coincide.
An intuition that comes out is that a few rare links will connect parts of the population that are different and otherwise largely disconnected, and this also reminds one of the concepts of \emph{weak ties} and \emph{structural holes} that we have introduced in Section \ref{sec:wattsstrogatz} (there the underlying distance between nodes was given by an exogenous Eudlidean topology, while here it is given by the exogenous characterization).
\cite{Jackson_book,jackson2008average} surveys properties of such models, and  in Section \ref{sec:implication_homophily}  we discuss \cite{golub2012homophily} who use a benchmark case  for their analysis.

\bigskip

In general, one could think that the characteristics of an individual is not just a single label, but that it is better described by a vector of (potentially continuous) variables: age, education, skills, political opinion, rate of self--identification with a particular culture\dots 
This representation may not be completely exogenous (as discussed above), but one can still treat it as such in the context of a model.
In this way every node $i$ is characterized a vector $\vec{\alpha}_i \in \mathbb{R}^m$, where $m$ is the number of characteristics that we consider, and we can think of $i$ as a point in an $m$--dimensional space, that can be endowed with many possible metrics.
In this setting, \cite{gauer2014continuous} and \cite{iijima2010social} are  two recent theoretical papers which assume that the probability of linking $p(i,j)$, of two nodes $i$ and $j$, depends in a monotonic decreasing way on their distance in this $m$--dimensional space. 
The two approaches differ in the distance metric they consider: \cite{gauer2014continuous} consider the Euclidean distance when $m=1$, while \cite{iijima2010social} adopts the \emph{$k$'th norms}, in which the distance between two points in the
type space is the $k$'th smallest distance among the $m$ dimension-wise distances between them.
The latter choice allows the expected realized network to obtain a higher level of clustering and maintain a small diameter.

\subsection{Linking based on friendship}
\label{sec:cjp}

We will discuss two friendship models that allow for an interesting analysis of homophily.  The first is based on a matching model and the second is based on a growing network formation model.  

Let us, thus,  return first to the analysis of \cite{currarini2009economic,currarini2010identifying}, which we anticipated in Section \ref{sec:pools} as a model of search and matching.
In their model each individual is endowed with a type that comes from a discrete set $\theta\in\Theta$ (we continue to denote the proportion of type-$\theta$ agents in the population by $w_{\theta}$), and the combination of random meeting and preference-based biases over types have some interesting consequences.
If a type has a high representation in the matching pool, and if members of that type place high value on meeting others of the same type, then the expected benefit of search will remain high for this type, and so they will have a greater number of total connections.  Building on this intuition, and assuming types prefer meeting others of the same type, if a given type constitutes a large share of the population, then they will optimally search more, and so their representation in the matching pool will be even higher than $w_{\theta}$, leading to an even more attractive calculation for their continued search.  Thus, the two sources of biases are inherently intertwined and, in this way, can reinforce each other.  

  Their model allows one to understand a number of empirical regularities of segregation patterns in friendship networks as a consequence of homophily-related biases.  
 The first empirical regularity is that individuals in larger groups tend to form more total connections.  Formally, the equilibrium number of connections should be increasing in $w_{\theta}$.  It is important to note that this prediction is not universally born out in the literature.  In the present model, a necessary condition for agents of different types to prefer different numbers of total connections is that they have type-based preferences.  Thus, in a given application, if all biases are opportunity-based, rather than preference-based, one would not expect such a pattern in the data.  

The second empirical regularity  relates to \emph{inbreeding homophily} $h_{\theta}$, 
which captures the excess proportion of same-type friends over the type's representation in the population, normalized to have a maximum value of unity.  The empirical regularity is that $h_{\theta}$ is inverse-U shaped, showing the greatest levels of homophily for groups of intermediate sizes, but generally positive even for the smallest and largest groups.  In the model, this prediction relies on a combination of preference-based on opportunity-based same-type biases.  The opportunity-based bias is important because it allows all groups, including the small groups, to meet same-types at relatively high rates.\footnote{%
While \cite{currarini2009economic} assume exogenous meeting biases, a  micro--foundation for this is provided by \cite{currarini2011simple}. \cite{currarini2011simple} assume that agents have a trade--off in making new connections between homophily and diversity.
This results in different searching strategies which in turn  endogenously biases their matching pools.}

We now turn attention to the analysis of \cite{bramoulle2012homophily}, who extend the model from \cite{jackson2007meeting} that we discussed in Section \ref{sec:JR} to incorporate types and homophily-based biases.  Agents are again assigned a type from a finite set.  Random meetings are biased in the following way.  Let $p(\theta,\theta')$ denote the probability that an entering agent of type $\theta$ meets an agent of type $\theta'$ in the random meeting process.  In general, $p(\theta,\theta')$ can differ arbitrarily from the type frequencies, allowing for biases.  Then, conditional on the realization of $p$, an agent from the appropriate type is drawn uniformly at random.\footnote{%
The search-based meetings are allowed to be similarly biased, although much of the analysis assumes neutrality for these meetings. }
Letting $p(\theta)$ be the probability of being born of type $\theta$ (the $w_{\theta}$ discussed above), we let $P_{j}^{t}(\theta_{t},\theta_{j})$ denote the probability that a
node born in period $j$ of type $\theta_{j}$ receives a link from a node of
type $\theta_{t}$ born at time $t>j$, the following expression is a
mean--field approximation of the overall linking probability, that generalizes equation (\ref{eq:JR}):
\begin{equation}
P_{j}^{t+1}(\theta,\theta_{j})=m_{R}\frac{p(\theta)p(\theta,\theta_{j})}{%
tp(\theta_{j})}+ m_{N}\sum_{\theta^{\prime}\in\Theta}p(\theta)p(\theta,%
\theta^{\prime})\frac{\sum_{\lambda=j}^{t}P_{j}^{\lambda}(\theta^{\prime},%
\theta_{j})}{tp(\theta^{\prime})}\frac{1}{m}  \ \  , \label{5}
\end{equation}
where $m$, $m_R$ and $m_n$ are those described in Section \ref{sec:JR}.
This multi--dimensional set of $|\Theta| \times |\Theta|$ differential equations can be expressed in a more compact way as
\begin{equation}
\label{eq:with_B}
\mathbf{P}_{j}^{t+1}=\frac{ m_{R}}{t}\mathbf{B} +\frac{m_{S}}{m t}%
\mathbf{B} \sum_{\lambda=j}^{t}\mathbf{P}_{j}^{\lambda}\ \ ,
\end{equation}
where
\begin{equation}
B (\theta,\theta^{\prime})\equiv p(\theta)\frac{p(\theta,\theta^{\prime})%
}{p(\theta^{\prime})}.
\label{eq:B}
\end{equation}
Following the mean field approach, we can transform equation (\ref{eq:with_B}) into a continuous differential equation in matrix form, which has a unique solution.

This model generates a number of testable predictions.  One finding of interest is that, for quite different reasons, inbreeding homophily exhibits the same inverse-U shape as a function of group size as in the \cite{currarini2009economic} analysis.  A main object of interest in the \cite{bramoulle2012homophily} analysis concerns \emph{integration}, i.e., the tendency of an agent's set of friends to become less and less biased over time.  In their linking process, the friends of friends meetings creates a channel through which an agent meets a less biased set of nodes compared to direct search of her immediate neighborhood.  Since as an agent ages, she tends to be found more and more through the friends of friends channel, compared to through random search, her neighborhood begins to reflect a diminished bias over time.  In particular, when search-based meetings are unbiased, the authors show that in the long run, every agent's local neighborhood converges to the type frequencies in the population.  Nonetheless, the network at large can still be heavily biased, as the neighborhoods of younger agents will generally be same-type biased.  

This finding is actually quite reminiscent of the work of \cite{chaney2014network} in the context of international trade.  Indeed, the main mechanism at work is similar in the two cases.  In the setup of \cite{chaney2014network}, the friends of friends channel translates into the finding that the trading partners of one's trading partners are located further away in space, and he finds an analogue to long run integration.  We refer to the reader to \cite{ChaneyChapter19} in this book for more discussion.


One takeaway from these papers is that a combination of type-based biases, along with their equilibrium consequences, is quite capable of capturing a broad range of empirically relevant homophilous patterns.  While the two models are quite different from each other, both of them impose a combination of biases that have interesting interactions that generate the main results.  Whether a search model or a network growth model is more appropriate for a given application, or whether one wants to explicitly model preferences, depends on the details and goals of a given study.

\subsection{Implications of homophily}
\label{sec:implication_homophily}



While the previous subsections have discussed some of the ways in which homophily can arise, we now turn our attention to understanding some of the various consequences of homophilous connection patterns.  That is: what is the importance of homophily in social networks?

In many important applications, the interactions between agents, the structure of which is captured through the network, are influenced by the types of the agents.  Agents with similar views or backgrounds may communicate at lower cost, for example.  Or it could be that agents with similar opinions are more likely to talk to each other, and this will clearly have an influence on how opinions evolve over time.  The spread of epidemics will be strongly influenced by how segregated different groups in the population are from each other.  

Taking up the first idea, it is possible to study opinion evolution under simple  rules of updating.\footnote{%
See \cite{GolubChapter12} in this book for a deeper discussion and for the distinction between Bayesian and na\"ive learning models.}   What will the presence of homophily imply about eventual opinions in a population?   \cite{golub2012homophily,golub2012network} study these questions and derive some interesting results.  Consider a process in which agents update their opinions by taking weighted averages of the opinions of their neighbors.  Such a model was first proposed by \cite{french1956formal} and \cite{degroot1974reaching}, and is a natural way to capture a na\"ive learning dynamic or a myopic best reply under a simple utility specification.  The key insight here is that homophily is tied to segregation, in that highly homophilous societies tend to have groups that live in relative isolation from each other, due to the relative sparsity of connections between groups.  That segregation, in turn, has important implications for opinion dynamics.  Under general conditions (essentially that the network is path-connected and aperiodic), opinions converge to a consensus in the long-run.  But an important object of interest is the speed of that convergence.  \cite{golub2012homophily,golub2012network} demonstrate that more homophilous societies converge much more slowly.  

Coming now to the idea of contagion in networks,
\cite{golub2012homophily,golub2012network}   show also that on an exogenous network homophily does not play a role in the contagion process, as long as the diameter of a network is kept fixed.
However in an economic process of contagion in which agents react endogenously to the risk of being infected,  \cite{galeotti2013strategic}  present a simple model that shows how the level of homophily impacts the contagion of an infectious disease.  In their work, there are two populations who interact with each other, with a parameter $\beta$ that controls the probability that any given interaction is same-type.  Thus, for $\beta>1/2$, the connections are homophilous, while for $\beta<1/2$, most connections are across groups as in, for example, the case of sexual contacts and gender.  Agents strategically choose whether or not to immunize at a cost.  Agents who remain vulnerable anticipate outcomes according to the steady-state of a standard SIS diffusion process, see, e.g. \cite{bailey1975mathematical}.  Under these dynamics, agents become infected through their connections to other infected agents, and they recover at a given exogenous rate. 

In equilibrium, agents balance the cost of immunization against the cost of time spent infected.  To illustrate the importance of homophily, consider starting from a world in which the two groups are completely separate, i.e., $\beta=1$ and there are no cross-group connections.  Suppose that the cost of immunization is slightly lower (relative to the benefits of being healthy) in, say, group $A$, so that group $A$ immunizes at a slightly higher rate.  Now consider the effects of gradually mixing the two groups, so that the world is homophilous, but not perfectly so.  The essence of the result is that equilibrium outcomes in the two groups diverge, so that the consequence of mild homophily is to magnify the slight underlying difference between the groups.  As individuals from group $A$ come into contact with those in group $B$, they are exposed to slightly more infection, since group $B$ has lower immunization.  This causes group $A$ to adjust upwards their already higher immunization rate.  Conversely, group $B$ benefits from the lower exposure due to interacting with group $A$, so that they have less incentive to immunize. As the two groups interact more and more, eventually group $B$ stops immunizing all together.  

One remark is that this is a very simple model, relying on a basic diffusion process and a simple homophily structure.  A promising avenue of further research is to explore various directions for modeling strategic decisions that depend on type-dependent contact patterns.  

Another domain where network structure and, in particular, homophily, will have important effects is in market settings including, for example, product adoption.  In a world in which consumers communicate with each other about product experiences, there are a range of pricing and competition questions to be addressed, so as to understand differences in such environments compared to cases in which consumers have complete information ex ante.  Two papers that study demand and pricing effects are  \cite{campbell2013word} and \cite{chuhay2014strategic}.  We think this is an area that will gain increasing attention.

\subsection{Empirical estimation of homophilous biases}


As pointed out in the beginning of this section, even when we are sure that the classification into types is exogeneous, it is very difficult to empirically disentangle the sources of observed segregation.
\cite{schelling1971dynamic} already demonstrated that this can depend on the choices of one or another group, or possibly many groups, and on the meeting constraints that people from different groups have, because, e.g., some variable (e.g.~workplace or hobbies) is correlated with type, and this bias in opportunities can itself recursively be caused by all of the former factors.
From the point of view of the econometrician,  homophilous biases can be included in one--shot homophilous models (Section \ref{sec:blockmodels}), matching pool models (Section \ref{sec:pools}), structural models (Section \ref{sec:structural}), or in some of the mean field results from growing network models (Section \ref{sec:growing}). In all cases the problem is to estimate the level of this bias for each group.

It is possible to analyze data where simultaneously more similar environments are observed, e.g.~many high schools where different ethnic groups or races are differently represented, as in the Add Health dataset.\footnote{%
\label{note:addhealth}
The National Longitudinal Survey of Adolescent Health (commonly referred to as Add
Health) is a program project that started in 1994 collecting data from a representative sample of almost 100 U.S.~High Schools, it was designed by J. Richard Udry, Peter S. Bearman, and Kathleen Mullan Harris.}
In this case the units of observation for the analysis are the average behavior of members of one type in each school, and it is clear that many environments are needed to obtain statistical significance.
Under these conditions, matching pool models are particularly suited, as they allow the analyst to specify also a bias in the meeting opportunities, as has been done by \cite{currarini2010identifying}, who find a great heterogeneity both in opportunities and in choices between the different ethnic groups.\footnote{%
Another possibility is to specify an underlying random network of opportunities on which agents establishes links with some preferential bias, as in 
\cite{franz2010observed}.}
Both sources of bias are significant in the data, and they also differ significantly across races.  For example, Asians and Blacks have a much stronger chance-based bias than Whites.  Estimated preference biases range from valuing inter-race connections at 90\% of the level of intra-race connections for Asians, to 55\% for Blacks. 

When instead the researcher has a single snapshot of the network, or data from only a limited number of points in time (as panel data on the same network, or only a few separated environments), then the neighborhood of each single node must be used as a unit of observation.
The most natural theoretical framework to test seems to be one--shot network formation models, and the benchmark models, where the probability of linking $p(i,j)$ depends only on the types of $i$ and $j$, have been applied to empirical estimation by \cite{newman2004detecting} and \cite{copic2009identifying} (in this case they are sometimes called \emph{block models}).
However, these applications have proved themselves not to be robust, and require Monte Carlo methods to estimate the confidence levels.
Another approach is to use structural models, as in \cite{mele2010structural}, \cite{goldsmith2013social} and \cite{graham2014}.
This approach provides more stable outcomes, but has the drawback of being computationally difficult, so that a modern computer is able to process only a network whose size is on the order of some hundreds of nodes.\footnote{Here we have highlighted only the models underlying the empirical approach on detecting homophily in social networks.
More details, and in particular the description of the methods, are in \cite{ChandrasekharChapter21} in this book.}

\section{Conclusion}
\label{sec:conclusion}


Thus far, the most successful models of network formation are based primarily on random events.  Analyses without an explicit random component produce severely stylized network structures.  As a result, our current theoretical toolkit for studying social networks in economics has benefited greatly from graph theory and combinatorial techniques in mathematics.  This approach has allowed us to learn a great deal about the connections between the micro-level stochastic processes of link formation and the large-scale structural characteristics they produce in a network.  Leveraging observational and empirical work on the structure of social networks to identify and calibrate models that best fit macroscopic network features, we can then make inferences about the linking processes by which those networks form.

The tradeoff inherent in this approach is that one need not model explicitly behavior as arising from incentives or objectives of the agents involved.  This omission largely precludes addressing welfare and policy questions.  For this reason, much of the random network formation literature has come from disciplines outside economics.  We think an area of particular interest to economists, and to those seeking to understand social networks, will be bringing together what is known about random network formation with analyses that bring economic behavior to the forefront.  For example, \cite{van2012risk} show how the ``friends of friends'' meeting process can originate from optimizing behavior of the agents under particular constraints. 

More generally, it is important to have a more complete understanding of the decisions that go into meeting and linking with other agents.  Some aspects of these events are random even from the perspective of the agents themselves.  Other events may be dictated by particular circumstances that are essentially deterministic to the agents, but are based on variables that are not observed by the analysts, and so appear to be random.  Thus, a complete picture must be informed by (i) an understanding of the properties of random formation processes, (ii) a set of theoretical frameworks with which to model the incentives of agents and understand their optimal behavior, and (iii) empirical work that identifies the relevant characteristics of agents and their environments, to bet understand their decisions, and (iv) finally structural work to estimate the resulting models.  


\subsection*{Acknowledgements}
\noindent
We gratefully acknowledge help from Ben Golub, Dunia L\'opez--Pintado, Angelo Mele, Yves Zenou, and participants at the 10th International Winter School on Inequality and Social Welfare Theory in Canazei.

\bibliographystyle{chicago}
\bibliography{pr_survey}
%
%
%

\enddocument